\documentclass[modern, times]{aastex701}

\usepackage{CJK}
\usepackage{makecell}

\makeatletter
\let\frontmatter@title@above=\relax
\makeatother

\begin{document}
\begin{CJK}{UTF8}{gbsn}  

\title{The reason for the occurrence of W-type contact binaries}

\author[orcid=0000-0003-2017-9151,gname=Jia, sname=Zhang]{Jia Zhang (张嘉)}
\affiliation{Yunnan Observatories, Chinese Academy of Sciences, Kunming 650216, China}
\email{yelangdadi@163.com}

\author[gname=Sheng-Bang, sname=Qian]{Sheng-Bang Qian (钱声帮)}
\affiliation{Department of Astronomy, Key Laboratory of Astroparticle Physics of Yunnan Province, Yunnan University, Kunming 650091, China}
\email{qiansb@ynu.edu.cn}

\correspondingauthor{Li-Ying Zhu (朱俐颖)}
\author[gname=Li-Ying, sname=Zhu]{Li-Ying Zhu (朱俐颖)}
\affiliation{Yunnan Observatories, Chinese Academy of Sciences, Kunming 650216, China}
\email[show]{zhuly@ynao.ac.cn}

\author[gname=Xu-Zhi, sname=Li]{Xu-Zhi Li (李旭志)}
\affiliation{School of Mathematics and Physics, Anqing Normal University, Anqing 246133, China}
\affiliation{Institute of Astronomy and Astrophysics, Anqing Normal University, Anqing 246133, China}
\email{Lixuzhi@ustc.edu.cn}

\begin{abstract}

For more than half a century, the puzzling W-type phenomenon in contact binaries has challenged astrophysicists. In these systems, the less massive component exhibits a higher surface temperature than its more massive companion, which is a reversal of the typical A-type configuration, where the more massive star is hotter. This counterintuitive temperature inversion defies the basic stellar physics and still lacks a widely accepted explanation. In this study, we assembled a sample of over 3,000 extensively observed contact binaries and derived their complete set of physical parameters. Our statistical analysis revealed a strong positive correlation between the occurrence of W-type contact binaries and the intensity and frequency of magnetic activities. This result strongly supports the hypothesis that magnetic activities are the primary driver of the W-type phenomenon and offers a compelling explanation for the observed transitions between the W-type and A-type.

\end{abstract}



\section{Introduction} \label{sec:Introduction}

Contact binaries are a unique class of stellar systems in which two stars share a common envelope \citep{1968ApJ...151.1123L, 1979ApJ...231..502L}. Although this configuration was once considered unstable \citep{1984QJRAS..25..405S}, observations revealed that contact binaries are remarkably abundant—numbering around 400,000 according to the VSX catalog \citep{2006SASS...25...47W}—and far outnumber other types of variable stars. These systems are notable for three key characteristics: (1) The component stars exhibit nearly identical surface temperatures despite often having significantly different masses \citep{2005ApJ...629.1055Y}. (2) A well-defined linear relationship exists between their orbital periods and surface temperatures, making them valuable as distance indicators \citep{1994PASP..106..462R, 2020RAA....20..163Q}. (3) The observed minimum mass ratio can be predicted theoretically \citep{2007MNRAS.377.1635A, 2009MNRAS.394..501A, 2024NatSR..1413011Z}.

Despite the near-equality of two component temperatures, subtle differences remain, prompting the traditional classification of contact binaries into A-type and W-type systems \citep{1970VA.....12..217B}. In A-type systems, the more massive star is hotter, whereas in W-type systems the more massive star is cooler.

Researchers have long debated the structural differences and evolutionary connections between these two types. Early work suggested that W-type systems have shallower common envelopes \citep{1973Ap&SS..22..381L} and tend to exhibit later spectral types with lower mass ratios \citep{1973A&A....25..249M}. However, subsequent observations have cast doubt on the idea that differences in common envelope thickness or mass ratios alone can account for the A-type/W-type distinction. Some studies proposed that A-type systems possess lower density and angular momentum, implying a more advanced evolutionary stage \citep{1981ApJ...245..650M,1988MNRAS.231..341H}. \citet{2001MNRAS.328..635Q} suggested that contact binaries may oscillate around a critical mass ratio of approximately 0.4. Moreover, data from over 100 contact binaries indicate that A-type systems generally have higher mass and angular momentum, arguing against a simple evolutionary progression from W-type to A-type \citep{2006MNRAS.370L..29G}. Recently, \citet{2020MNRAS.492.4112Z} analyzed a sample of 117 contact binaries and found that W-type and A-type systems exhibit different radial density distributions, indicating that they have different evolutionary origins.

\citet{2024ApJ...975..231X} found that eight contact binaries exhibit transitions between A-type and W-type. Among these, four (V839 Oph, AC Boo, V1848 Ori, and V694 Peg) transitioned from A-type to W-type, two (AM Leo and AH Cnc) transitioned from W-type to A-type, and two (RZ Com and FG Hya) underwent two transitions (either from A-type to W-type and back to A-type, or vice versa). These binaries span a wide range of parameters: mass ratios from 0.11 to 0.76, primary star masses from 0.7 to 1.7 $M_{\odot}$, and fill-out factors $f$ from 0.06 to 0.91. Their thickness of common envelope varies in a large range, demonstrating that the transition phenomenon can occur across a broad parameter space, rather than being confined to a special narrow range. \citet{2016RAA....16..157P} also provided two examples of the W to A transition (EM Psc and V1191 Cyg). We will show that the phenomenon of A/W-type transitions serves as independent evidence for the conclusions presented in this paper.

The ``near-equality of temperatures in contact binaries is still mysterious" \citep{2005ApJ...629.1055Y}, and the particularly perplexing observation that the less massive star in W-type systems is hotter, called W-phenomenon \citep{1970VA.....12..217B, 1984QJRAS..25..405S}, presents an even greater challenge. Energy transfer from the hotter star can at best equalize the temperatures of the two components, but it cannot reverse the temperature ratio. Although many studies have explored the distinctions and potential evolutionary links between A-type and W-type systems, the underlying cause of the W-phenomenon remains unresolved.

In this paper, we employ a large sample of contact binaries, combined with extensive physical parameters and statistical methods, to investigate the widespread occurrence of the W-type phenomenon and reveals magnetic activities as the underlying cause.

\section{Photometrically Analyzed Contact Binaries}

The sample used in this study comprises contact binaries that have undergone light curve analysis to derive their relative parameters, which are essential for obtaining their full physical properties. Although the total number of known contact binaries is very large (approximately 0.4 million according to the VSX catalog \citealp{2012yCat....102027W}), only a limited number have been analyzed through light curve modeling.

Our sample is assembled from four sources: 380 systems analyzed by \citet{2020PASJ...72..103L} using \textit{Kepler} data, 318 systems analyzed by \citet{2023MNRAS.525.4596D} using \textit{TESS} data, 2335 systems analyzed by \citet{2020ApJS..247...50S} using data from the Northern Catalina Sky Survey, and 688 individually studied systems compiled by \citet{2021ApJS..254...10L}, yielding a total of 3721 contact binaries. We cross-matched these 3721 binaries with the Simbad database and found that 16 systems had no corresponding entries, while 249 binaries appeared multiple times across different catalogs. After removing duplicates and unmatched sources, we obtained a final sample of 3580 contact binaries.

All these binaries have parameters derived from light curve analysis, including orbital period, mass ratio, temperature ratio, radius ratio, the ratio of radius to semi-major axis, and inclination angle. Additionally, the catalogs provided by \citet{2021ApJS..254...10L} and \citet{2020PASJ...72..103L} include an important flag indicating whether starspots were incorporated in the light curve modeling. Since starspots serve as direct evidence of the existence of magnetic activities, this information is a key component in our argument that magnetic activities underlies the W-type phenomenon.

\section{Cross-matching contact binaries with \textit{LAMOST} and \textit{Gaia} for atmospheric parameters}

We cross-matched the collected 3,580 contact binaries with the \textit{LAMOST} DR11 \citep{2015RAA....15.1095L} and \textit{Gaia} DR3 catalogs \citep{2016A&A...595A...1G,2023A&A...674A...1G}. The coordinate uncertainties for these binaries are all within 0.3 arcseconds, based on the major axes of the error ellipses provided by the Simbad database \citep{2000A&AS..143....9W}. We adopted a cross-matching radius of 0.5 arcseconds for \textit{Gaia} and 1 arcsecond for \textit{LAMOST}.

The choice of cross-matching radius directly influences the accuracy of the matching process: a smaller radius reduces the probability of false matches but may also decrease the number of successfully matched sources. After careful consideration, a radius of 0.5 arcseconds was deemed optimal for \textit{Gaia}, ensuring a very low false match rate (less than 1\%) while minimizing missing sources. For \textit{LAMOST}, a matching radius of 1 arcsecond is appropriate given the lower target density and the spatial resolution limitations of its telescope.

In this paper, all atmospheric parameters from \textit{Gaia} DR3 were obtained from BP/RP spectra, specifically through the GSP-Phot model. While \textit{Gaia} offers several models for atmospheric parameters—including GSP-Spec (General Stellar Parametrizer from Spectroscopy), ESP-HS (Extended Stellar Parametrizer for Hot Stars), and ESP-UCD (Extended Stellar Parametrizer for Ultra Cool Dwarfs)—only GSP-Phot and GSP-Spec are capable of providing all three atmospheric parameters (\texttt{Teff}, \texttt{logg}, and \texttt{[Fe/H]}). When considering the number of targets, GSP-Phot significantly outperforms GSP-Spec, supplying atmospheric parameters for 3128 targets in this study compared to just 21 from GSP-Spec. Consequently, GSP-Phot was the model ultimately selected.

When a target possesses atmospheric parameters from both \textit{LAMOST} and \textit{Gaia}, we prioritize \textit{LAMOST}'s parameters, primarily because \textit{LAMOST}'s resolution is significantly higher than BP/RP. Ultimately, 1234 targets in this paper have atmospheric parameters sourced from \textit{LAMOST}, while 2055 targets utilize data from \textit{Gaia} DR3 GSP-Phot.

\section{Calculating the absolute parameters}

\subsection{Calculating absolute stellar parameters from stellar atmospheric parameters using Isochrones database}\label{sec:cal_absolute_stellar_parameters}

Light curve analysis can provide only the relative parameters of a binary system. To determine the absolute parameters, it is necessary to know either the mass or the radius of at least one component.

For binary stars, two primary approaches exist for obtaining absolute parameters. The first method relies on stellar evolutionary models, using atmospheric parameters to calculate fundamental stellar properties such as mass and radius. The second method utilizes radial velocity curves, which, when combined with light curve analysis, allow for the direct determination of stellar masses.

Each method has its advantages and limitations. In this study, we adopt the first approach by utilizing isochrone databases generated from stellar evolutionary models. By interpolating spectroscopically derived atmospheric parameters, we obtain fundamental stellar properties, which enables the determination of a large number of parameters for statistical analysis. In contrast, the second method, which depends on radial velocity measurements, is highly demanding observationally and therefore does not provide large-scale data for statistical studies.

Two stellar evolutionary databases were used in this study. The first is \textit{PARSEC} (PAdova and TRieste Stellar Evolution Code; \citealp{2012MNRAS.427..127B, 2014MNRAS.444.2525C, 2015MNRAS.452.1068C, 2014MNRAS.445.4287T, 2017ApJ...835...77M,2019MNRAS.485.5666P,2020MNRAS.498.3283P}), version CMD 3.6. We specifically adopted a Reimers mass-loss coefficient ($\eta_{Reimers}$) of 0.2 on the Red Giant Branch and a two-part power-law Initial Mass Function (IMF) from \citet{2001MNRAS.322..231K, 2002Sci...295...82K}, and \citet{2013pss5.book..115K}. We used an isochrone database covering a broad range of metallicities ([M/H] from -2.19 to +0.7 with a step of 0.02) and ages (log(age/yr) from 6.6 to 10.13 with a step of 0.01). A total of 22 million sets of stellar parameters were obtained to interpolate the atmospheric parameters of the targets.

The second evolutionary database is \textit{MIST} (MESA Isochrones \& Stellar Tracks; \citealp{2016ApJS..222....8D, 2016ApJ...823..102C, 2011ApJS..192....3P, 2013ApJS..208....4P, 2015ApJS..220...15P, 2018ApJS..234...34P}). \textit{MIST} constructs isochrones based on Modules for Experiments in Stellar Astrophysics (MESA), a widely used open-source evolution code. \textit{MIST} covers a broad range of ages (5 $\le$ log(age/yr) $\le$ 10.3), masses (0.1 $\le$ $M/M_{\odot}$ $\le$ 300), and metallicities (-4.0 $\le$ [Z/H] $\le$ 0.5). The \textit{MIST} database we used accounts for stellar rotation, with the rotational velocity uniformly set to $V/V_{crit} = 0.4$, where $V$ is the equatorial rotation velocity, and $V_{crit}$ is the critical rotation velocity (above which the star would begin to lose mass from its surface or break apart). A mature Python package called \texttt{Isochrones} \citep{2015ascl.soft03010M}, version 2.1, was used to interpolate stellar properties by employing the \textit{MIST} database.

Using isochrone interpolation, we derived the mass and radius for the primary components of contact binaries; however, stellar age was not determined. While significant mass transfer during contact binary formation profoundly alters a star's evolutionary path and physical parameters, single-star models can still provide instantaneous parameters like mass and radius based on current atmospheric data. They cannot, however, yield the star's age, as its evolutionary history deviates fundamentally from that of a single star.

Existing research has compared the differences between calculation results from binary star models that account for mass transfer and those from single-star models. For two binary systems, KIC 10736223 and OO Dra, \citet{2020ApJ...895..136C, 2021ApJ...920...76C} determined the parameters of their pulsating stars. This was accomplished using both single-star and binary-star models, with the latter incorporating mass accretion and stellar rotation. Despite these pulsating stars having accrued 2 and 1.7 times their initial mass, respectively, over the course of their binary evolution, the outcomes from both modeling approaches were remarkably similar. Specifically, the mass discrepancies between the single-star and binary models were minor, at merely 2.5\% and 1\%. Similarly, radius deviations were also minimal, registering at 0.7\% and 0.3\%. This suggests that even significant mass accretion does not fundamentally alter a star's internal and external structure such that it no longer aligns with a single-star paradigm, allowing for accurate mass and radius determination via single-star models. However, it's crucial to note that single-star models are inherently limited to calculating a star's current state and cannot reconstruct its past evolutionary trajectory.

For the factor of rotation, the uniform rotational velocity of $V/V_{crit} = 0.4$ by \textit{MIST} is certainly not identical to those of the stars in our sample, it generally exceeds the actual rotational velocities. We compared the results from \textit{PARSEC} (which does not include rotation) and \textit{MIST} (which does); the comparison is shown in Figure \ref{fig:Compar_parsec_and_mist} in Appendix \ref{appendix:Compar_mass_for_methods_models_inputpars}. The median relative deviation between the two is only 2.7\%, which is minor compared to the discrepancies shown in Figure \ref{fig:Compar_stellar_model_and_orbital_dynamical} (mass comparison with dynamical measurements). Therefore, while stellar rotation does influence the results, its impact is minor for our sample.

In addition to the fundamental stellar parameters, we also needed to calculate two other physical quantities for subsequent statistical research: the relative thickness of the common envelope and the Rossby Number. We calculated the primary star's radius in the pole direction (perpendicular to the binary orbital plane) and the radius of the primary star's Roche lobe in the pole direction (i.e., the corresponding radius at the critical equipotential surface). The difference between these two radii defines the thickness of the common envelope ($Th_{CE}$). The physical quantity we needed for statistical study is the ratio of the common envelope's thickness ($Th_{CE}$) to the primary star's radius ($R_1$) (in the pole direction), referred to as the relative thickness of the common envelope ($Th_{CE}/R_1$). The second physical quantity is the Rossby Number, defined as $R_o = P_{\text{rot}}/\tau$, where $P_{\text{rot}}$ is the stellar rotation period (the orbital period for contact binaries) and $\tau$ ($log\tau = 1.16 - 1.49*logM_1 - 0.54*logM_1^{2}$) is the convective turnover time \citep{2011ApJ...743...48W}. These two parameters will be statistically analyzed as physical quantities characterizing the strength of magnetic activity.

\subsection{The applicability of the single-star model to the component star of contact binaries}

We derived the fundamental stellar parameters using atmospheric parameters obtained from spectroscopic observations. Although the observed spectrum is a composite of both components, we assume that it is primarily dominated by the more luminous component star, whose surface properties are best reflected in the measurements. In most cases, the more luminous component is also the more massive star.

In contact binaries, the two components typically exhibit similar surface temperatures due to heat transport within the common envelope. Their surface metallicities should also be nearly identical due to differential rotation and convection. The primary distinguishing factor between the two components may be their surface gravity (\texttt{logg}). Therefore, the deviation in treating the observed spectrum as that of the primary star may arises from differences in \texttt{logg}.

To study the discrepancies between the spectroscopically derived \texttt{logg} and the true \texttt{logg} of the primary star, we adopted an alternative interpolation scheme for isochrones. Instead of using \texttt{logg}, we interpolated stellar density, which can be directly obtained from light curve analysis \citep{2017MNRAS.466.1118Z}. A comparison of results between the new approach (using temperature, density, and metallicity) and the traditional approach (using temperature, \texttt{logg}, and metallicity) is presented in Figure \ref{fig:Compar_TRhoF1_and_T1gF} in Appendix \ref{appendix:Compar_mass_for_methods_models_inputpars}. The results indicate no systematic bias between the two methods, suggesting that \texttt{logg} derived from spectroscopy statistically roughly represents the true \texttt{logg} of the primary star. To maintain consistency between spectroscopic and photometric calculations, all statistical analyses in this study are based on the temperature, density, and metallicity approach.

Although we used stellar evolutionary models to estimate the fundamental parameters of the primary star, this approach is not suitable for the secondary star. Stellar evolutionary models are designed for single stars evolving in isolation, free from strong external interactions. In contact binaries, the primary star dominates the system's luminosity, and its surface layers remain relatively undisturbed by the presence of the secondary star. In contrast, the secondary star undergoes significant modifications due to mass exchange and tidal interactions, causing its properties to deviate substantially from those of an isolated star. Our calculations indicate that estimating the secondary's mass based on its atmospheric parameters would result in severe overestimation. Since binary interactions profoundly alter the secondary's evolutionary state, traditional classifications such as main sequence or post-main sequence are no longer applicable. Consequently, we only determine the fundamental parameters of the primary star. The mass and radius of the secondary star are then obtained using the mass ratio and radius ratio derived from light curve analysis.

The dominant source of error are mainly from the contact configuration of the binary system itself, rather than from mass transfer during binary evolution, or the stellar rotation.

Figure \ref{fig:Compar_stellar_model_and_orbital_dynamical} in Appendix \ref{appendix:Compar_mass_for_methods_models_inputpars} compares the masses calculated in this paper with those derived by previous methods using radial velocities, showing median relative deviations of 20\% and 27\%. We contend that the primary source of this deviation stems from the mutual influence between component stars within the contact configuration. Eliminating this effect would necessitate knowledge of the primary star's surface atmospheric parameters in a detached state, which is beyond the current capabilities of theoretical modeling. Presently, no existing model can simulate contact binaries (while current models can account for mass transfer and stellar rotation) because the three-dimensional structure of the common envelope fundamentally exceeds the one-dimensional approximation inherent in current stellar models. Despite these significant parameter errors, they remain valuable for the statistical investigations conducted in this work. The following section will demonstrate that these parameters, although with large uncertainties, can still yield reliable statistical conclusions.

\section{Results and Discussion}

We compiled a sample of 3,580 contact binaries from \citet{2020PASJ...72..103L}, \citet{2020ApJS..247...50S}, \citet{2021ApJS..254...10L}, and \citet{2023MNRAS.525.4596D}. All of these systems have been analyzed via light curve modeling to derive key parameters such as orbital period, mass ratio, radius ratio, and temperature ratio. Among these, 3,234 sources have atmospheric parameters—effective temperature (\texttt{Teff}), surface gravity (\texttt{logg}), and metallicity (\texttt{[Fe/H]})—obtained from \textit{LAMOST} or \textit{Gaia} observations, while 135 systems possess radial velocity curves for both components.

Using both the atmospheric parameters and the light curve derived quantities, we calculated the physical parameters for each component, including mass, radius, and luminosity. The contact binary catalog is described in Appendix \ref{appendix:catalog}.

Our statistical analyses reveal that magnetic activities is the key driver behind the formation of W-type contact binaries. In comparison to A-type systems, the primary star (i.e., the more massive component) in W-type contact binaries tends to have a lower mass, and exhibits stronger or frequent magnetic activities.

Furthermore, the extensive sample enabled us to identify several intriguing phenomena, including a positive correlation between primary mass and metallicity, the evolutionary direction towards lower mass ratio, and non-coplanarity between binary orbits and the Galactic disk.

\subsection{Magnetic Activities as the Driver of W-type Contact Binaries}

W-type contact binaries are considered anomalous because the less massive component exhibits a higher surface temperature, despite their significant mass differences (sometimes by factors of 5 to 10). Our statistics show that W-type systems are more common than A-type ones, with a ratio of approximately 1.4 (2100 W-types to 1463 A-types), highlighting the prevalence of this anomaly.

One might attribute the higher temperature of the lower-mass component to an unusual evolutionary phase. For instance, if the star were in a post-AGB stage, its surface temperature could exceed that of a more massive main-sequence star. However, such extreme evolutionary cases are rare and cannot account for the widespread occurrence of W-type contact binaries, particularly in systems with significant mass difference.

Given the constraints imposed by the cosmic age and the fact that both stars in a binary system are formed simultaneously, any special or extreme evolutionary scenario for the low-mass star cannot explain the commonality of W-type systems, though it may account for individual special cases.

Regardless of whether a contact binary is classified as A-type or W-type, the two components generally exhibit nearly identical surface temperatures—a result of the highly efficient heat transport within the common envelope. Differential rotation, rather than convection, is considered the more likely and effective mechanism for this energy transfer \citep{2005ApJ...629.1055Y}, rapidly smoothing out any initial temperature disparities.

Thus, to explain the prevalent W-type phenomenon, it is crucial to maintain the observed near-equality of surface temperatures. The difference between W-type anomalies and A-type normalcy lies in the subtle temperature variations that emerge under the condition of overall thermal uniformity.

\subsubsection{Magnetic Activities as a Trigger for the W-type Phenomenon}



Magnetic activities on a stellar surface can modify the local and overall surface temperature. Phenomena such as dark spots, faculae, flares, prominences, and coronae introduce localized temperature increase or decrease, thereby affecting the star's average temperature. If magnetic activities lead to a decrease of primary temperature or an increase of secondary temperature, it could potentially cause the W-type phenomenon, where the less massive star appears hotter than its companion.

While various magnetic activity phenomena can influence a star's surface temperature, if we only consider those with a significant and prolonged impact (several weeks or more) in the optical band, then typically only dark spots and the frequently co-occurring faculae are relevant. Flares or coronal mass ejections, though potentially violent, have very short timescales (less than one day) and primarily enhance brightness in the ultraviolet and X-ray bands. Other magnetic activity phenomena have negligible effects on the optical band. As a result, the observed temperature variations in contact binaries are predominantly attributed to dark spots and faculae. Referring to the Sun, a higher incidence of cool spots leads to a higher incidence of hotter faculae, which more than compensates for the cooler spots making the Sun actually brighter during sunspot maximum. If spots appear on the surface of the secondary component of a contact binary, it is likely to also cause an increase in its average temperature, further leading to the W-type phenomenon.

In light curve analysis, ``dark spots'' are modeled as low-temperature, circular regions introduced to reproduce asymmetries in the observed brightness variations. While the modeled spots provide evidence for magnetic activities, they represent an averaged effect rather than the locations of actual starspots. Either component of the binary may host spots, and fitting a spot on either star can yield a satisfactory match to the data. Thus, the inclusion of a spot in the model confirms the presence of surface temperature inhomogeneities without specifying their exact distribution.

Even though the overall temperature variation induced by magnetic activities is modest, it is significant when considering that the temperature differences between the two components of a contact binary are typically small. Consequently, the temperature changes from magnetic activities can sufficiently lower the temperature of the more massive star or raise that of the less massive star, thereby producing the W-type phenomenon. While magnetic activities can also produce A-type systems, such effects are considered part of normal behavior, whereas the W-type phenomenon remains an anomalous signature that this study aims to explain.

\subsubsection{The correlation between W-type occurrence frequency and magnetic activities}

Theoretically, magnetic activities can induce a W-type phenomenon. Here, we demonstrate that there is a strong positive correlation between magnetic activities and the probability of observing W-type contact binaries—that is, the stronger or more frequent the magnetic activities, the higher the likelihood of a system to be the W-type.

To quantify this relationship, we employ several indicators of magnetic activities. Common proxies include the X-ray-to-bolometric luminosity ratio, flare frequency, magnetic flux, and Rossby Number. In our study, we utilize four indicators: the frequency of starspot occurrence ($N_{\text{Spot}}/N_{\text{No Spot}}$), stellar mass ($M_1$), relative common envelope thickness ($Th_{CE}/R_1$), and Rossby Number ($R_o$). Among these, the probability of starspot occurrence positive reflects magnetic activities, while the other three indicators show strong negative correlations with it (i.e., lower stellar mass, thinner relative common envelopes thickness, or smaller Rossby Numbers correspond to stronger or frequent magnetic activities).

\begin{figure*}[h]
\fig{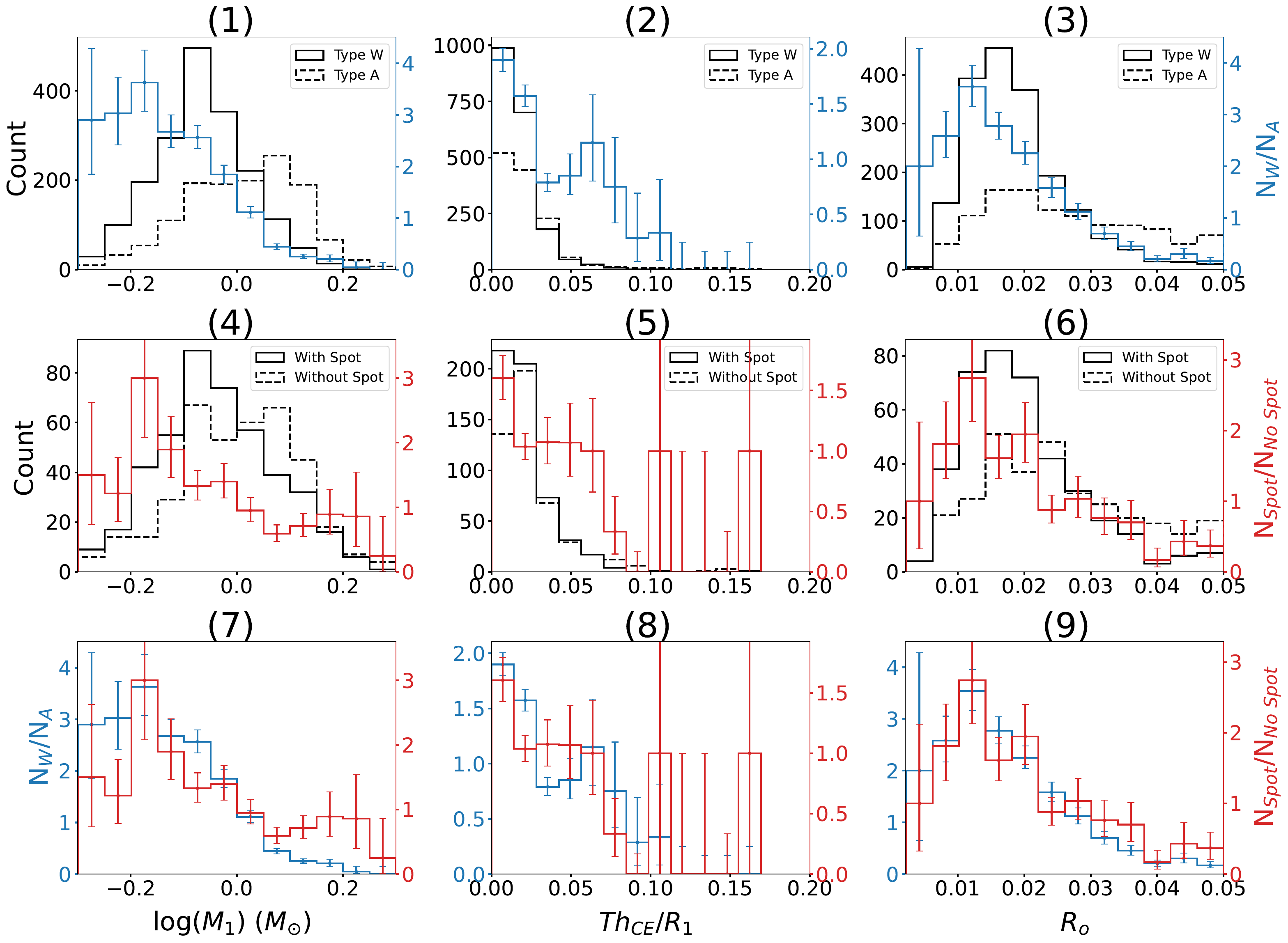}{1\textwidth}{}
\caption{Panels 1-3: The number of W-type ($N_W$), A-type ($N_A$), and their ratio $N_W/N_A$ as a function of the primary star mass ($\log M_1$), common envelope thickness ($Th_{CE}/R_1$), and Rossby Number ($R_o$). Panels 4-6: Same as Panels 1-3, but for the number of contact binaries with starspots ($N_{\text{Spot}}$) and without starspots ($N_{\text{No Spot}}$). Panels 7-9: Comparison of $N_W/N_A$ and $N_{\text{Spot}}/N_{\text{No Spot}}$ from Panels 1-6. The color of the curves corresponds to the color of the Y-axes. \label{fig:dist_WA_spots}} \end{figure*}

Panels 1-3 of Figure \ref{fig:dist_WA_spots} present the distributions of three parameters: the logarithm of the primary star mass ($\log M_1$), the common envelope thickness ($Th_{CE}/R_1$), and the Rossby Number ($R_o$). In these panels, the numbers of W-type ($N_W$) and A-type ($N_A$) systems are shown separately, while the blue curves represent the ratio $N_W/N_A$, serving as a proxy for the probability of W-type occurrence as a function of these parameters.


Panels 4-6 display the distributions of the same three parameters, but the data are classified based on the presence of starspots—$N_{\text{Spot}}$ represents the number of binaries with starspots, while $N_{\text{No Spot}}$ denotes those without. The red curves represent the ratio $N_{\text{Spot}}/N_{\text{No Spot}}$, which serves as an indicator of the probability of starspot occurrence as a function of these parameters. Note that only \citet{2020PASJ...72..103L} (private communication) and \citet{2021ApJS..254...10L} provided starspot data, whereas \citet{2020ApJS..247...50S} and \citet{2023MNRAS.525.4596D} did not; thus, the number of contact binaries with starspot indicators (1021) is smaller than the total sample (3580). Panels 7-9 overlay the blue and red curves to directly compare $N_W/N_A$ and $N_{\text{Spot}}/N_{\text{No Spot}}$.

The data in Panels 1-3 indicate that as the primary star mass, common envelope thickness, and Rossby Number increase, the ratio $N_W/N_A$ generally decreases. Similarly, Panels 4-6 show that $N_{\text{Spot}}/N_{\text{No Spot}}$ also decreases with these parameters. Panels 7-9 further demonstrate a strong synchronization between $N_W/N_A$ and $N_{\text{Spot}}/N_{\text{No Spot}}$, indicating that decrease in one ratio correspond to decrease in the other.

The occurrence of starspots causes surface temperature inhomogeneities, leading to asymmetries in the light curves. In light curve analysis, the inclusion of a spot directly indicates the presence of magnetic activities on the stellar surface. The strong correlation between starspot occurrence and the W-type frequency (as shown in Panels 7-9) directly supports the conclusion that magnetic activities plays a key role in the emergence of W-type contact binaries.

A general consensus is that there is an anti-correlation between stellar mass and magnetic activities \citep{2009ARA&A..47..333D, 2012LRSP....9....1R}, meaning that lower-mass stars exhibit more frequent and intense magnetic activities. If a spot (along with faculae) appears on the secondary star, it may increase its overall average temperature, thereby favoring the formation of a W-type configuration. Consequently, systems with larger masses tend to exhibit weaker magnetic activities and lower $N_W/N_A$ ratios, as supported by the trends observed in Panels 1 and 4.

\citet{2003MNRAS.342.1260Q} argued that the relative common envelope suppresses magnetic activities, with stronger magnetic activities occurring when the envelope is thinner. Based on ROSAT X-ray observations, \citet{2001A&A...370..157S} demonstrated that the X-ray flux from contact stars is 4-5 times lower than that from rapidly rotating single stars. Similarly, \citet{2001AJ....121.1084H} noted that short-period detached binaries, such as ER Vul and UV Leo, exhibit significantly higher levels of starspot and chromospheric activities before evolving into contact binaries. These findings suggest that a thicker common envelope diminishes magnetic activities, thereby reducing the $N_W/N_A$ ratio, as evidenced by the data in Panels 2 and 5.

The Rossby Number $R_o$ is a critical parameter linking stellar rotation, convection, and magnetic activities. As $R_o$ decreases, both the relative flare luminosity \citep{2016ApJ...829...23D} and the X-ray-to-bolometric luminosity ratio \citep{2011ApJ...743...48W} increase, indicating enhanced magnetic activities. Our calculations of $R_o$ for each contact binary (Panels 3 and 6) show that as $R_o$ decrease, both $N_W/N_A$ and $N_{\text{Spot}}/N_{\text{No Spot}}$ increase.

We note two issues that were omitted from the previous discussion for clarity: (1) The two leftmost bins in the distributions of stellar mass and $R_o$ in Figure \ref{fig:dist_WA_spots} show a decrease that deviates from the overall negative correlation, and (2) the correlation statistics for the secondary star mass are not presented.

Regarding the first issue, we believe there may be two possible causes: 1) measurement errors. The error bars in Figure \ref{fig:dist_WA_spots} are solely Poisson errors derived from the statistical counts and do not consider the inherent uncertainties in the parameters. Figure \ref{fig:WA_distribution_by_false_mass} in Appendix \ref{appendix:random_check_for_uncertainty_of_mass} demonstrates the impact of mass errors on the statistical results (described in Section \ref{sec:uncertainties_and_impact}). When mass errors are taken into account, the two lower bins on the far left disappear in the vast majority of the random test results, resulting in a monotonic decreasing distribution. 2) saturation of the stellar magnetic field. Below a certain $R_o$ value, magnetic saturation occurs, and parameters such as magnetic flux \citep{2009ApJ...692..538R}, relative flare luminosity \citep{2016ApJ...829...23D}, and the X-ray-to-bolometric luminosity ratio \citep{2011ApJ...743...48W} cease to increase and may plateau or even decrease. In contact binaries, the saturation level of $R_o$ (approximately 0.01) is about an order of magnitude lower than that for single stars (approximately 0.1). As noted earlier, the X-ray flux in contact binaries is 4-5 times weaker than in rapidly rotating single stars \citep{2001A&A...370..157S}, implying that $R_o$ in contact binaries may be much lower at saturation than in single stars.

Regarding the second issue, the correlation between the secondary star mass and $N_W/N_A$ (presented in Appendix \ref{appendix:relation_for_m2}) is weak. The surfaces of both component stars in a contact binary are enveloped by a common envelope. The atmospheric conditions within this common envelope should primarily be determined by the primary star. Energy transfer within the common envelope, facilitated by differential rotation, can efficiently transfer both mass and energy. The secondary star's surface temperature being significantly higher relative to its mass also indicates that the common envelope's temperature is mainly dictated by the primary star. Therefore, the secondary star's mass does not play a decisive role in its own surface temperature, and it is highly probable that the secondary star's mass also has a weak influence on surface magnetic activity. Consequently, the correlation between the secondary star's mass and the occurrence of the W-type phenomenon is weak.

In summary, the trends observed in primary star mass, relative common envelope thickness, and Rossby Number—as well as the starspot occurrence ratio $N_{\text{Spot}}/N_{\text{No Spot}}$—strongly support the conclusion that enhanced magnetic activities increases the probability of the W-type phenomenon.

\subsubsection{The uncertainties in primary mass and their impact on statistical results}\label{sec:uncertainties_and_impact}

The stellar parameters in this study were derived using stellar evolutionary models. To assess the reliability of these parameters, the most robust and independent validation method is direct comparison with values obtained through independent method, that is the orbital dynamical measurement. Unlike model-based calculation, orbital dynamical measurements rely solely on observational constraints—specifically, the radial velocity curves of both components, combined with the binary's orbital period and inclination—to determine stellar masses.

Among the 3,580 contact binaries in our sample, we cross-matched them with SB9 (The Ninth Catalogue of Spectroscopic Binary Orbits; \citealp{2004AandA...424..727P}) and identified 135 systems with double-lined radial velocity measurements. Of these, 99 systems also have atmospheric parameters from \textit{LAMOST} or \textit{Gaia}, enabling a direct comparison between model-based and model-free masses. This comparison is shown in Figure \ref{fig:Compar_stellar_model_and_orbital_dynamical} in Appendix \ref{appendix:Compar_mass_for_methods_models_inputpars}, where stellar masses derived from the \textit{MIST} database are plotted against their corresponding dynamical measurements.

The relative deviations span a wide range, from 0.6\% to 739\%, with a median deviation of 27\% for the Temperature-Density-Metallicity interpolation and 20\% for the Temperature-\texttt{logg}-Metallicity interpolation. A natural question arises: could such large uncertainties affect our main statistical conclusions—particularly, the observed anti-correlations between $N_W/N_A$ and $N_{\text{Spot}}/N_{\text{No Spot}}$ with primary mass?

To address this concern, we conducted a Monte Carlo simulation. Using the distribution of relative deviations shown in Figure \ref{fig:Compar_stellar_model_and_orbital_dynamical}, we applied these deviations to all primary masses to generate artificial (simulated) mass values. Specifically, each primary mass was multiplied by a random relative deviation (from Figure \ref{fig:Compar_stellar_model_and_orbital_dynamical}) to generate an artificial mass value. The distribution of the randomly selected relative errors is identical to that shown in Figure \ref{fig:Compar_stellar_model_and_orbital_dynamical}.

To mitigate the effects of statistical fluctuations, we generated 5,000 sets of simulated primary masses, each containing the same number as the observed primary masses (3,234). For each set of simulated masses, we recalculated the distributions as shown in Panels 1 and 4 of Figure \ref{fig:dist_WA_spots}. The ensemble of these 5,000 distributions is displayed in Figure \ref{fig:WA_distribution_by_false_mass} in Appendix \ref{appendix:random_check_for_uncertainty_of_mass}. The results demonstrate that after accounting for the big uncertainties of masses, the anti-correlations between $N_W/N_A$ and $N_{\text{Spot}}/N_{\text{No Spot}}$ with primary mass remain statistically valid. Thus, the uncertainties illustrated in Figure \ref{fig:Compar_stellar_model_and_orbital_dynamical} do not compromise the key trends shown in Figure \ref{fig:dist_WA_spots}.


\subsubsection{The randomness and variability of magnetic activities}

The previous sections have demonstrated that magnetic activities can induce the W-type phenomenon, supported by four distinct observational evidences. In this section, we emphasize that magnetic activities is inherently random and variable. On the surface of a given star, magnetic activities can be intermittent, vary in intensity, and manifest in various forms (e.g., dark spots, flares, prominences, and coronae).

This inherent randomness leads to two key consequences: (1) Not all contact binaries that exhibit magnetic activities are W-type. (2) A single contact binary may switch between W-type and A-type over time due to the variability of magnetic activities. 

Both of the two consequences are consistent with observations. Clearly, not all contact binaries are W-type. The highest proportion of W-type systems occurs when the primary star's mass is between 0.63 and 0.71 $M_{\odot}$, reaching 78\% (as shown by the third and highest bar in panel 1 of Figure \ref{fig:dist_WA_spots}). For the second consequence, as we introduced in the Introduction section, observations have found eight contact binaries exhibiting A-type/W-type transitions, including two systems that underwent two times of transition. This phenomenon precisely corroborates the randomness and variability of magnetic activities.

If the differences between A-type and W-type systems were solely rooted in the internal stellar structure, such structures would not change significantly over timescales of days to years, and the binary type would remain fixed. The observed rapid transitions, however, strongly suggest that the inherent randomness and variability of magnetic activities is the only plausible explanation for the switching between A-type and W-type phenomena.

Thus, while the presence of magnetic activities increases the probability of a contact binary exhibiting the W-type configuration, it does not guarantee a permanent classification. A W-type binary may subsequently transition to an A-type, and vice versa, reflecting the dynamic nature of magnetic activities.

\subsection{The distribution of spots from light curve analysis}

\begin{figure*}[h]
\fig{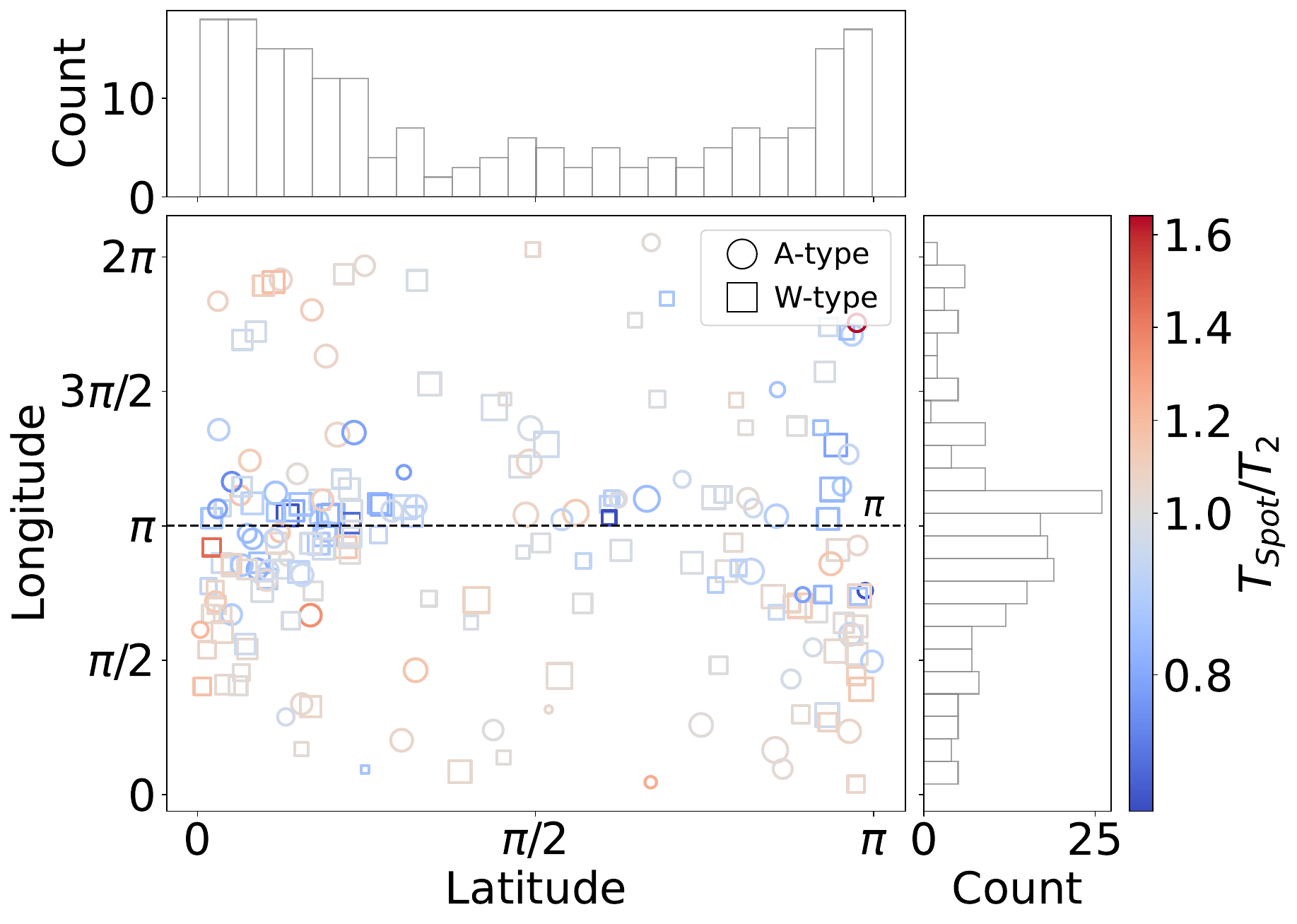}{0.9\textwidth}{}
\caption{Distribution of the spots from light curve analysis by \citet{2020PASJ...72..103L}. The point size represents the spot's area, and the color, defined by the colorbar, indicates the temperature ratio of the spot to the surrounding photosphere. The symbol shape denotes the binary type. The marginal histograms at the top and right correspond to the distributions along the horizontal and vertical axes, respectively.} \label{fig:spots_distribution}
\end{figure*}

To better understand the influence of spots, we analyzed the complete set of spot parameters from \citet{2020PASJ...72..103L}, which we obtained via private communication. The Figure \ref{fig:spots_distribution} shows the distribution of these spots on the stellar surface, detailing their temperature, size, and location.


It's crucial to distinguish the ``spots'' derived from light curve modeling from discrete, solar-like sunspots. Instead of representing individual spots, they are large-scale approximations of the inconsistent surface temperature. These modeled regions represent a net temperature inhomogeneity, effectively reproducing observed light curve asymmetries that arise from the collective effect of numerous smaller, unresolved features like spots and faculae.

Consequently, the absence of a spot in a model does not imply a lack of magnetic activity. It may simply indicate that the activity is distributed uniformly enough across the visible hemisphere to produce a symmetric light curve, making a localized spot parameter unnecessary.

\citet{2020PASJ...72..103L} placed all modeled spots on the secondary component, but this does not imply spots are physically absent on the primary component. In practice, fitting a spot on either component can often yield an equally good solution. The choice to place spots on the secondary is often pragmatic, based on the assumption that the cooler secondary might be more prone to spot activity and that such models frequently converge more readily.

The spot parameters from light curve models reveals several key characteristics: (1) While spots can appear anywhere, they show a statistical preference for locations near latitude 0 or $\pi$ and longitude $\pi$, corresponds to the polar regions, oriented towards the observer during eclipse. (2) Spots can be either cooler or hotter than the surrounding photosphere. (3) Neither the temperature nor the location of the spots shows any strict correspondence with the binary's A/W-type classification.

The spot properties—their location, size, and temperature—show no causal link to a binary's A/W-type classification. In the W-type contact binary stars, spots with the similar locations can be either cooler or hotter. Therefore, a single spot model cannot explain all W-type phenomena. We think that spots and the W-type phenomenon are independent consequences of underlying magnetic activity.

Furthermore, a substantial number of both W-type and A-type systems show no spots in their light curve solutions. In our sample, 227 W-type binaries were modeled without spots; for at least 74 of these, the authors deemed spots unnecessary due to the high symmetry of the light curves.

Given the significant impact large spots can have, we propose a refinement to how stellar temperatures are reported from light curve analysis. Conventionally, the reported temperature is that of the surface excluding the spot area. While this is a long-standing practice, it can be misleading when a spot covers a large fraction of the surface.

We recommend instead using the spot-corrected, global average temperature, as it provides a more physically meaningful characterization of the star's overall thermal state. Reclassifying systems using this global temperature would change the A/W type for approximately 15\% of the binaries in our sample, though the total counts for each type would remain stable due to reciprocal changes.

\subsection{The relationship between primary mass and metallicity}

\begin{figure*}[h]
\fig{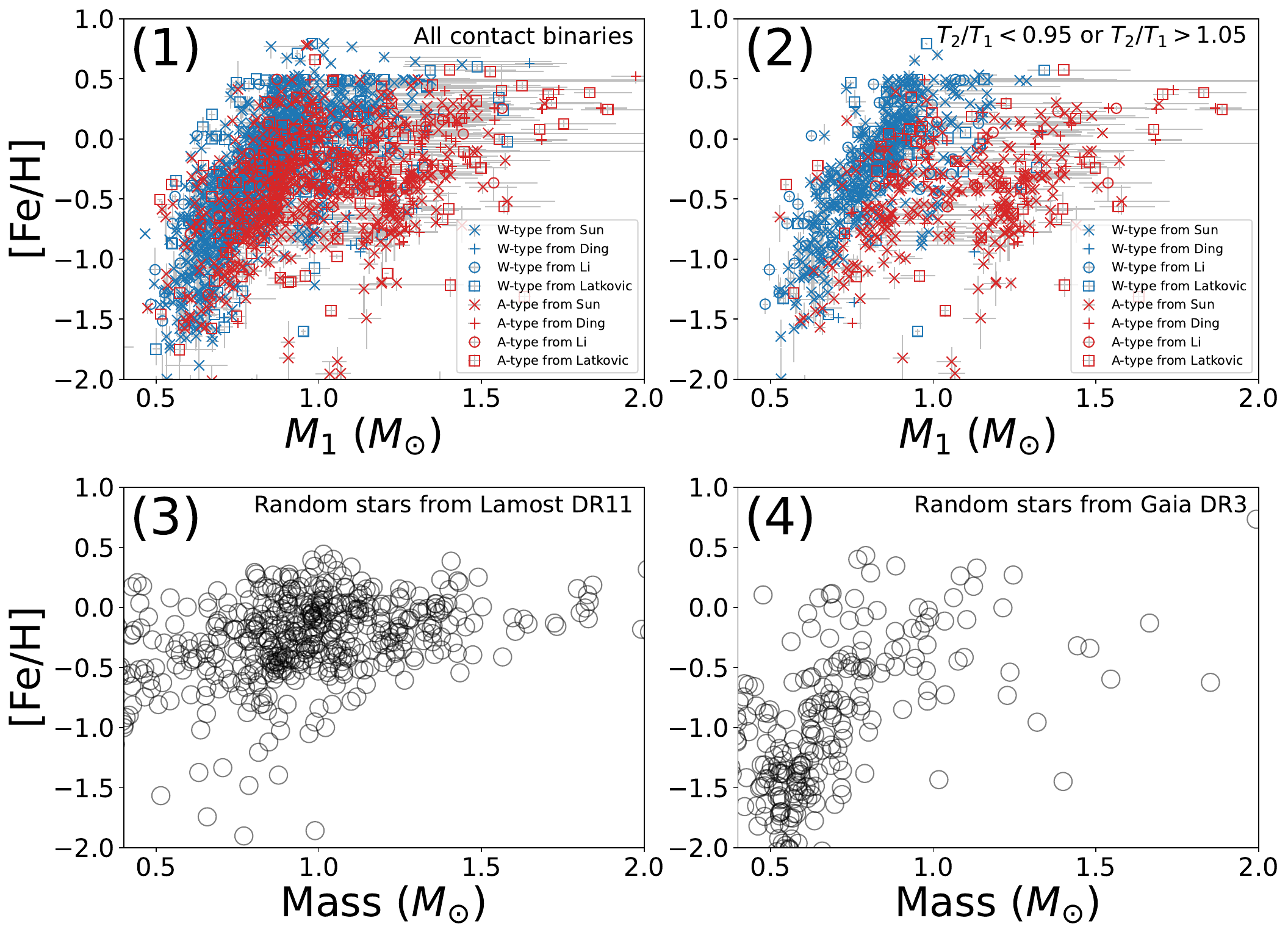}{1\textwidth}{}
\caption{The relationship between primary star mass \(M_1\) and metallicity [Fe/H] for all contact binaries (Panel 1), for contact binaries with a temperature difference between the two stars greater than 5\% (Panel 2). Different colors represent different binary types (red for A and blue for W), while different shapes indicate different data sources. The same relationship for 500 random stars from \textit{LAMOST} DR11 (Panel 3) and 300 random stars from \textit{Gaia} DR3 (Panel 4). \label{fig:mass_feh_relationship}} \end{figure*}

Figure \ref{fig:mass_feh_relationship} displays a clear positive correlation between primary star mass and metallicity, and it shows that A-type and W-type contact binaries occupy distinct regions in this parameter space. 



It should be noted that Panels 1 and 2 do not display every data point. A few outliers with excessively high mass or extremely low metallicity were omitted to improve the clarity of the distribution of majority points.

In Panel 2, to exclude targets with nearly identical temperatures for the two components (and thus ambiguous A- and W-type characteristics), only contact binaries with a temperature difference greater than 5\% are plotted. As a result, the red and blue points are more clearly separated, with blue W-types and red A-types both exhibiting a positive correlation—yet for a given primary mass, W-types show higher surface metallicity. Moreover, the distribution of W-types is more compact and the positive correlation is stronger compared to A-types.


The data for the contact binaries shown in Figure \ref{fig:mass_feh_relationship} are derived from both \textit{LAMOST} and \textit{Gaia} surveys. Utilizing data from \textit{LAMOST} or \textit{Gaia} alone yields similar results. Furthermore, to provide a comparison with other stars, we randomly selected 500 and 300 stars from the \textit{LAMOST} and \textit{Gaia}, respectively. We then determined their masses using the identical methodology and plotted them in Panels 3 and 4. The axes of these plots are scaled identically to the corresponding contact binary plots. The \textit{Gaia} data clearly shows a positive correlation, whereas the correlation is not as obvious in the \textit{LAMOST} data. This suggests that the mass-metallicity correlation observed in contact binaries might not be unique to them but may instead be a general characteristic of all stars. However, the correlation is distinctly stronger in the contact binaries, especially for the W-type systems.

\subsection{Distribution of mass ratio and the evolutionary direction of contact binaries}

The mass ratio distribution is subject to strong selection effects. Fortunately, the high photometric precision of \textit{Kepler} enables the detection of even weak planetary transit signals, ensuring that low-mass-ratio contact binaries are reliably identified. Consequently, we consider the mass ratio distribution derived from \textit{Kepler} data to be the least affected by selection biases.

\begin{figure*}[h]
\fig{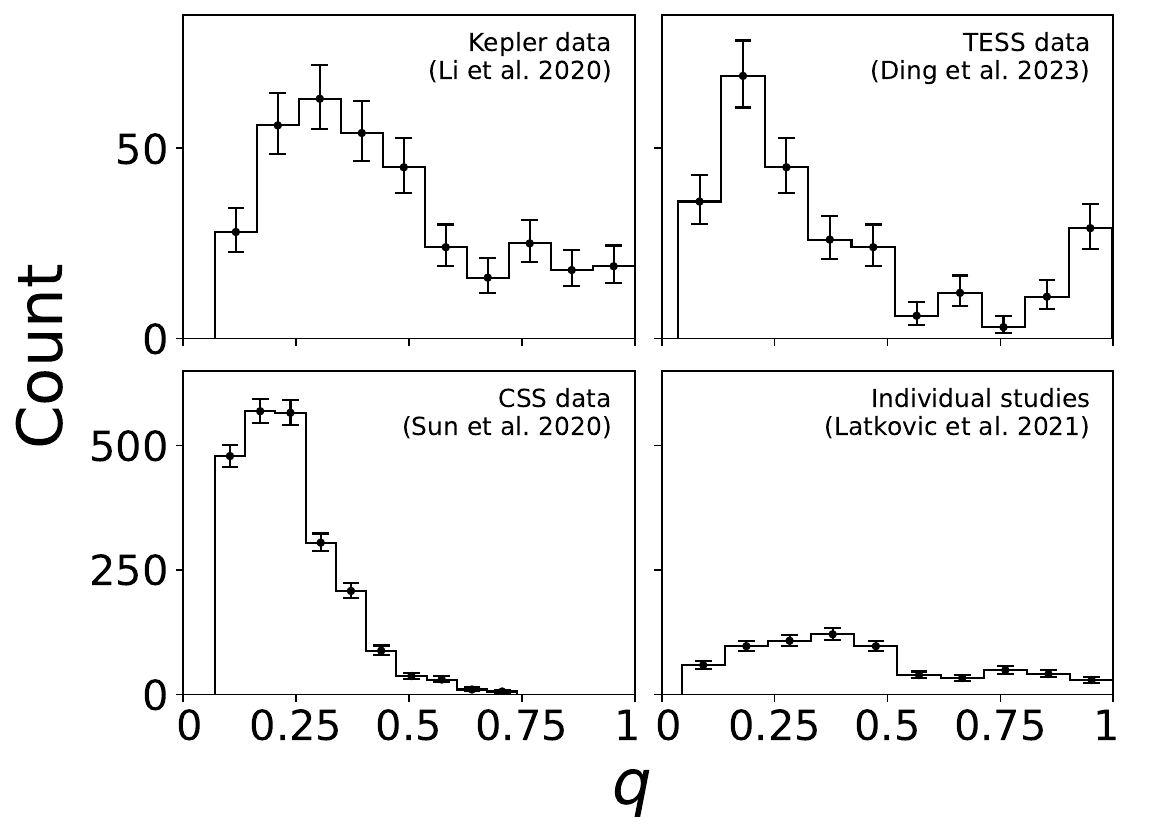}{0.8\textwidth}{}
\caption{Distribution of the mass ratio $q$. Data were sourced from \textit{Kepler} (top left), \textit{TESS} (top right), CSS (bottom left), and individual studies (bottom right). \label{fig:dist_q}}
\end{figure*}

Figure \ref{fig:dist_q} presents the distribution of the mass ratio $q$. In the \textit{Kepler} dataset (upper left panel), a prominent peak is observed at approximately $q\sim0.3$, with no corresponding peak near $q=1$. It is important to note that this distribution pertains to contact binaries, which, due to mass transfer processes, undergo a reversal in their mass ratios; hence, their current mass ratios are significantly different from their initial values.

Although the four distributions differ in detail, they all share a key feature: a pronounced peak at the low mass-ratio end. In the CSS dataset (lower left Panel), a high peak occurs closest to the lower end, accompanied by a clear cutoff at lower boundary. We hypothesize that this behavior may indicate the evolutionary direction of contact binaries toward lower mass ratios. This behavior becomes more intuitive if we check the relationship between mass ratio and common envelope thickness.

\begin{figure*}[h]
\fig{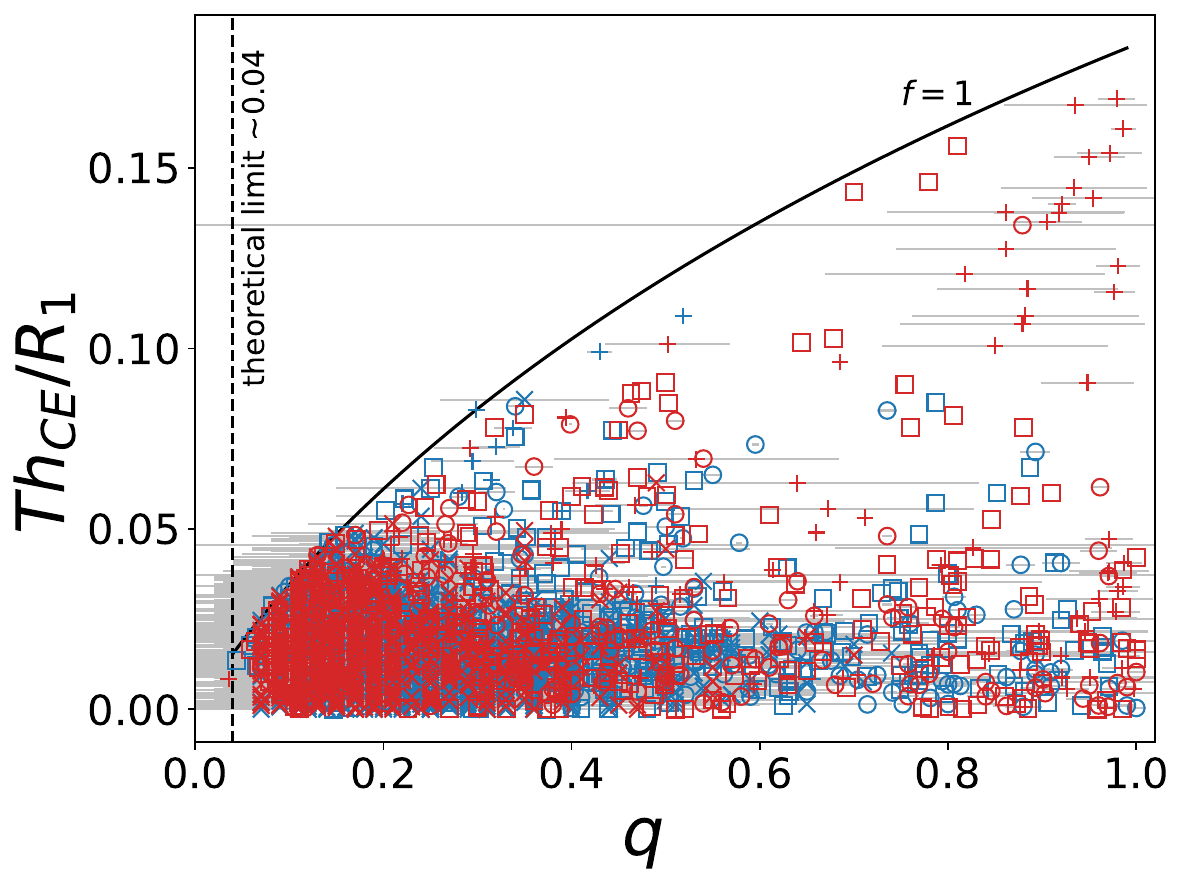}{0.8\textwidth}{}
\caption{The relationship between mass ratio $q$ and relative common envelope thickness $Th_{CE}/R_1$. The curve represents the boundary for a fill-out factor $f=1$. The data point colors and shapes follow the same scheme as in Figure \ref{fig:mass_feh_relationship}.} \label{fig:q_CEth2R1_relationship}
\end{figure*}

We derived the relative thickness $Th_{CE}/R_1$ of common envelope from the mass ratio $q$ and the fill-out factor $f$, both obtained via light curve analysis. Figure \ref{fig:q_CEth2R1_relationship} shows that, for different mass ratios, the relative thickness is bounded by different upper limits. Notably, all contact binaries fall below the black line representing $f=1$, and only systems with high mass ratios can potentially exhibit high $Th_{CE}/R_1$ values. In particular, when $Th_{CE}/R_1 > 0.1$, almost all systems are A-type, consistent with the findings in Panels 2, 5, and 8 of Figure \ref{fig:dist_WA_spots}. Furthermore, the lower limit of the mass ratio is close to the theoretical minimum of $\sim$0.04 \citep{2024NatSR..1413011Z}.

A question worth considering is why most data points are concentrated in the side of low mass ratio? In other words, why does a peak in number appear at the low mass-ratio end in Figure \ref{fig:dist_q}? One plausible explanation is that low-mass-ratio contact binaries is an evolutionary direction. As a contact binary evolves toward merging into a single star, its mass ratio continuously decreases toward zero. The sharp truncation observed at the lower mass ratio limit, rather than a gradual decline in number density, supports the idea that once $q$ falls below a certain theoretical threshold, the system is more likely to merge than to persist as a contact binary. 

The two binaries with the lowest observed mass ratios, TIC 125958765 and V1187 Her, have mass ratios of 0.036 and 0.044, respectively, and exhibit high fill-out factors (0.6 and 0.8), making them typical deep-contact binaries with extremely low mass ratios. We anticipate that such systems, as potential progenitors of luminous red novae, will undergo a merger event (e.g., an outburst) in the near future.

\newpage
\appendix

\renewcommand\thefigure{Appendix\arabic{figure}}    
\setcounter{figure}{0}

\section{Comparison of Primary Star Mass Across Different Methods, Evolutionary Models, and Input Parameters} \label{appendix:Compar_mass_for_methods_models_inputpars}

\begin{figure*}[h]
\fig{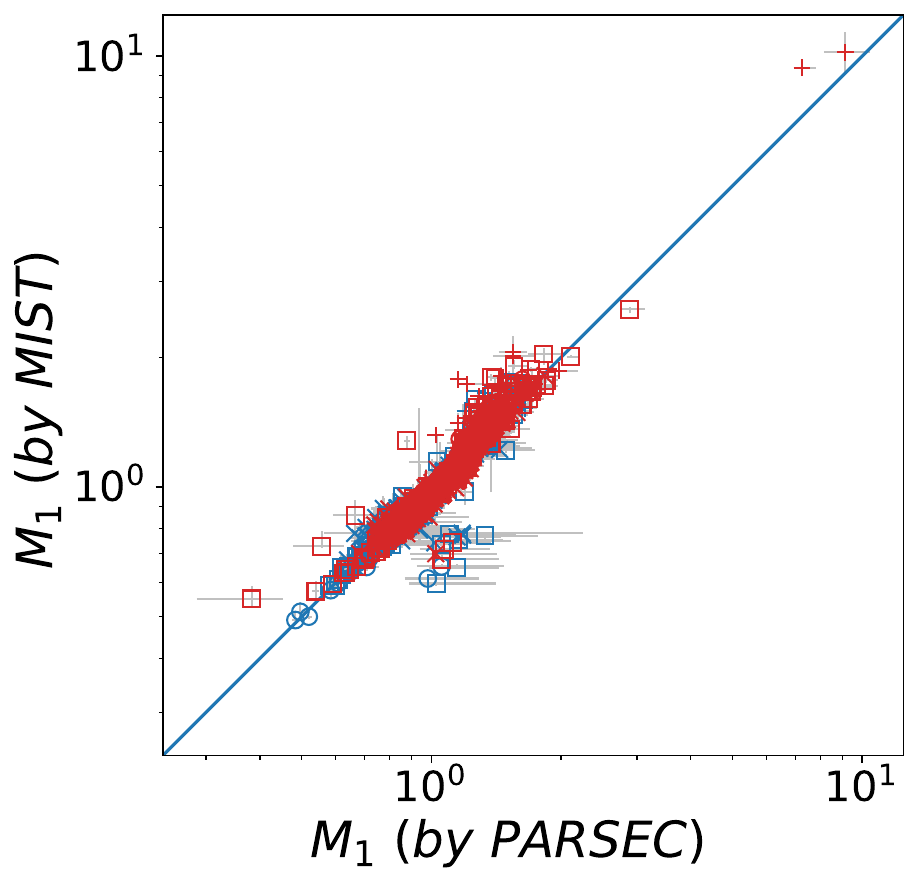}{0.8\textwidth}{}
\caption{Comparison of primary star mass between \textit{PARSEC} and \textit{MIST}. The data point colors and shapes follow the same scheme as Figure \ref{fig:mass_feh_relationship}. \label{fig:Compar_parsec_and_mist}}
\end{figure*}

Figure \ref{fig:Compar_parsec_and_mist} compares the primary star mass obtained from two stellar evolutionary models, \textit{PARSEC} and \textit{MIST}, using Temperature-Density-Metallicity as input parameters. The results from both models exhibit strong consistency, with a median relative deviation of only 3.8\%.

\begin{figure*}[h]
\gridline{\fig{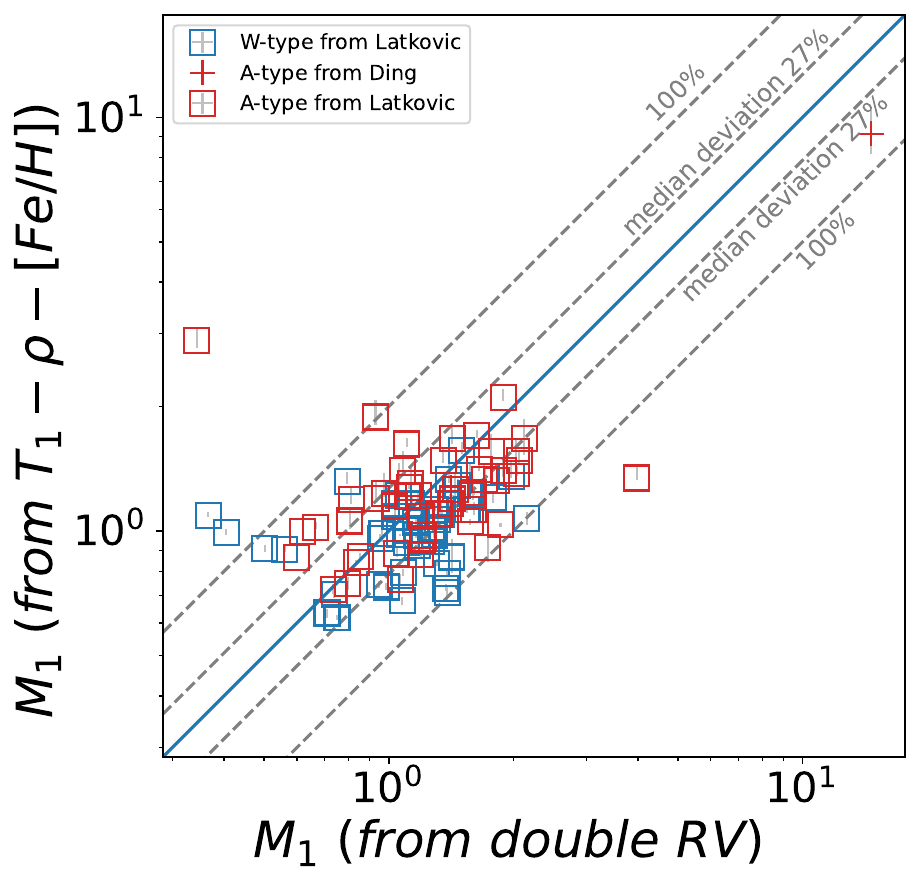}{0.45\textwidth}{\raisebox{1.5ex}{(1)}}
          \fig{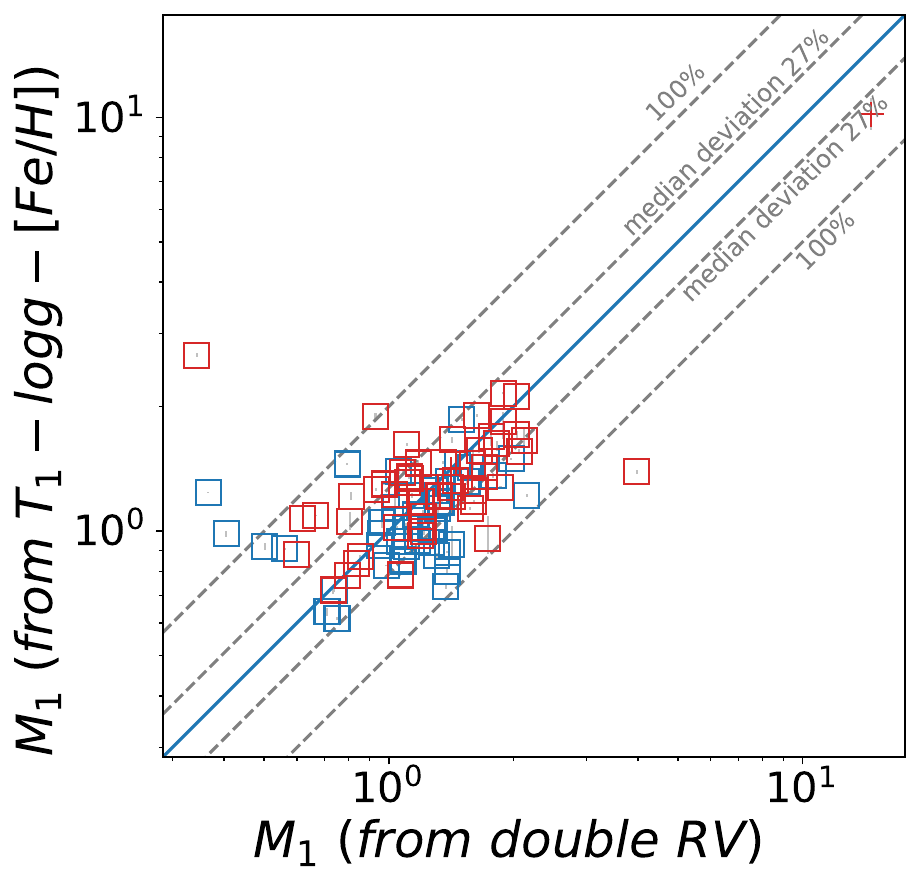}{0.45\textwidth}{\raisebox{1.5ex}{(2)}}
         }
\caption{Comparison of primary star mass from the \textit{MIST} stellar evolutionary model and orbital dynamical measurements (derived from radial velocity, inclination, and period). The input parameters are Temperature-Density-Metallicity (Panel 1) and Temperature-\text{logg}-Metallicity (Panel 2). The data point colors and shapes follow the same scheme as Figure \ref{fig:mass_feh_relationship}. \label{fig:Compar_stellar_model_and_orbital_dynamical}}
\end{figure*}

Figure \ref{fig:Compar_stellar_model_and_orbital_dynamical} compares the primary star mass derived from stellar evolutionary models and dynamical radial velocity (RV) measurements. The relative deviation ranges from 0.6\% to 739\% with a median value of 27\% when using Temperature-Density-Metallicity as input parameters, and from 0.07\% to 676\% with a median value of 20\% when using Temperature-\text{logg}-Metallicity.

\begin{figure*}[h]
\fig{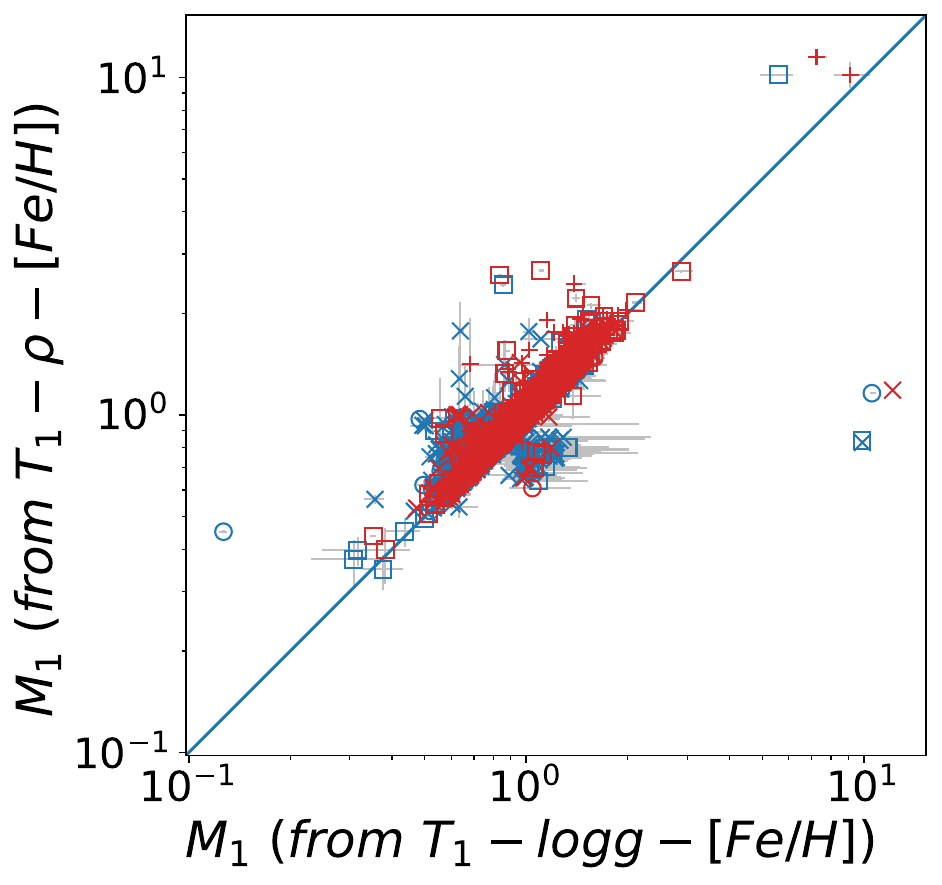}{0.8\textwidth}{}
\caption{Comparison of primary star mass obtained using $T_1 - \rho - [Fe/H]$ and $T_1 - \log g - [Fe/H]$ as input parameters within the \textit{MIST} model. The data point colors and shapes follow the same scheme as Figure \ref{fig:mass_feh_relationship}. \label{fig:Compar_TRhoF1_and_T1gF}}
\end{figure*}

Figure \ref{fig:Compar_TRhoF1_and_T1gF} examines the differences in primary star mass estimates when using two different sets of input parameters ($T_1 - \rho - [Fe/H]$ and $T_1 - \log g - [Fe/H]$), both by the \textit{MIST} database. The median relative deviation between these two approaches is 2.7\%, further validating the robustness of mass calculation against different parameter choices.

\section{Contact Binary Catalog} \label{appendix:catalog}

In addition to the four primary sources that provide light curve analysis results for contact binaries (\citealp{2020PASJ...72..103L}, \citealp{2020ApJS..247...50S}, \citealp{2021ApJS..254...10L}, and \citealp{2023MNRAS.525.4596D}), we incorporated four supplementary datasets and computed four additional sets of stellar parameters. A summary of these datasets is presented in Table \ref{tab:cat_statistics}.

This is currently the largest catalog of contact binaries containing a complete set of physical parameters. The full catalog is available for download at \url{https://astrophysics.cc/contact-binary-catalog/} or \url{https://zenodo.org/records/17385333}, along with the detailed parameter description files. Given that the catalog contains an extensive set of 476 parameters, of which only a subset is directly relevant to this study, we also provide a reduced table that includes only the 44 parameters pertinent to our analysis.

\centerwidetable  
\begin{deluxetable*}{l|l|l|l|l}
\tablecaption{Catalog Source Statistics}  \label{tab:cat_statistics}
\tabletypesize{\scriptsize}
\tablehead{
\colhead{Catalog Name} & \colhead{Reference}  &  \colhead{Main Parameters}  &  \colhead{  \makecell[c]{Number of\\ Parameters}} &  \makecell[c]{Number of \\ Targets}} 
\startdata  
\makecell[l]{Contact Binaries from \\ \textit{Kepler}} & \citealp{2020PASJ...72..103L} &  \makecell[l]{Orbital period, inclination, \\ mass ratio, temperature ratio, \\ spots}   & 28 & 380 \\
\hline
\makecell[l]{Contact Binaries from \\ \textit{CSS}} & \citealp{2020ApJS..247...50S} &  \makecell[l]{Orbital period, inclination, \\ mass ratio, temperature ratio}   & 28 & 2335 \\
\hline
\makecell[l]{Contact Binaries from \\ Individual Studies} & \citealp{2021ApJS..254...10L} &  \makecell[l]{Orbital period, inclination, \\ mass ratio, temperature ratio, \\ spots}   & 29 & 688 \\
\hline
\makecell[l]{Contact Binaries from \\ \textit{TESS}} & \citealp{2023MNRAS.525.4596D} &  \makecell[l]{Orbital period, inclination, \\ mass ratio, temperature ratio}   & 24 & 318 \\
\hline
\makecell[l]{The Ninth Catalogue of \\ Spectroscopic Binary Orbits}  &  \citealp{2004AandA...424..727P}  &  \makecell[l]{Radial velocity amplitude}   & 47 & 140 \\
\hline
\makecell[l]{Astrophysical Parameters \\ from Gaia DR3}  &  \citealp{2023AandA...674A..26C}  &  \makecell[l]{Atmospheric parameters: \\ \text{T$_{\rm eff}$}, \text{log g}, and \text{[Fe/H]}}   & 216 & 3128 \\
\hline
\makecell[l]{Cluster Membership \\ Determination in Gaia DR3}  &  \citealp{2023AandA...675A..68V}  &  \makecell[l]{Cluster name, \\ probability of membership}   & 18 & 20 \\
\hline
\makecell[l]{Atmospheric Parameters \\ from LAMOST DR11}  &  \citealp{2015RAA....15.1095L}  &  \makecell[l]{Atmospheric parameters: \\ \text{T$_{\rm eff}$}, \text{log g}, and \text{[Fe/H]}}   & 25 & 1234 \\
\hline
\makecell[l]{Parameters Calculated from \\ \text{T$_{\rm eff}$}, $\rho$, and [Fe/H] by MIST}  &  This paper  &  \makecell[l]{Primary mass, primary radius}   & 33 & 3234 \\
\hline
\makecell[l]{Parameters Calculated from \\ \text{T$_{\rm eff}$}, log g, and [Fe/H] by MIST}  &  This paper  &  \makecell[l]{Primary mass, primary radius}   & 33 & 3289 \\
\hline
\makecell[l]{Parameters Calculated from \\ \text{T$_{\rm eff}$}, $\rho$, and [Fe/H] by PARSEC}  &  This paper  &  \makecell[l]{Primary mass, primary radius}   & 36 & 2125 \\
\hline
\makecell[l]{Parameters \\ Calculated Manually \\ Based on the Above Catalogs}  &  This paper  &  \makecell[l]{Secondary mass, \\ unified mass ratio, \\ unified temperature ratio}   & 38 & 3580 \\
\hline
Total  &  ---  &  ---   & 476 & 3580 \\
\enddata
\end{deluxetable*}


\section{\texorpdfstring{The Distribution of $N_W/N_A$ and $N_{\text{Spot}}/N_{\text{No Spot}}$ from Monte Carlo testing}{The Distribution of NW/NA and N\_Spot/N\_No\_Spot from Monte Carlo testing}} \label{appendix:random_check_for_uncertainty_of_mass}

\begin{figure*}[h]
\gridline{\fig{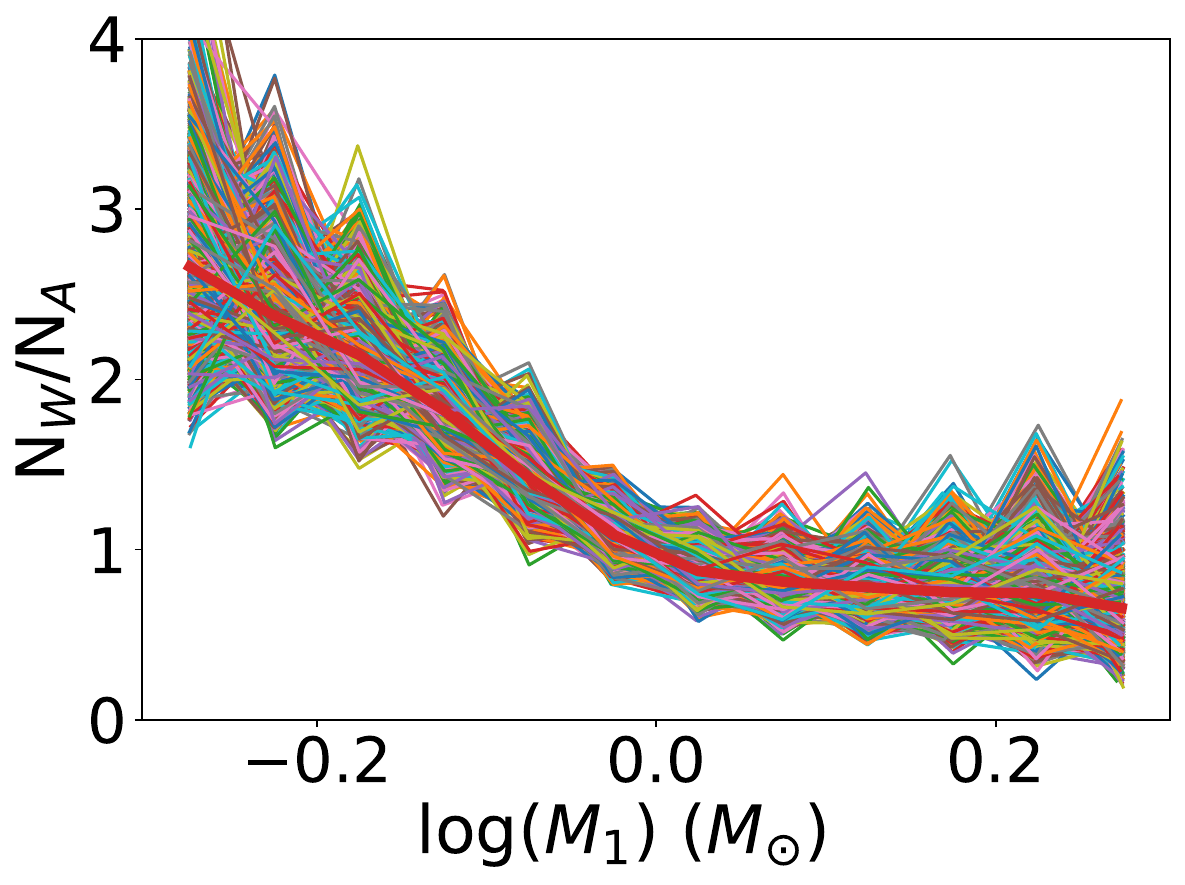}{0.48\textwidth}{\raisebox{1.5ex}{(1)}}
          \fig{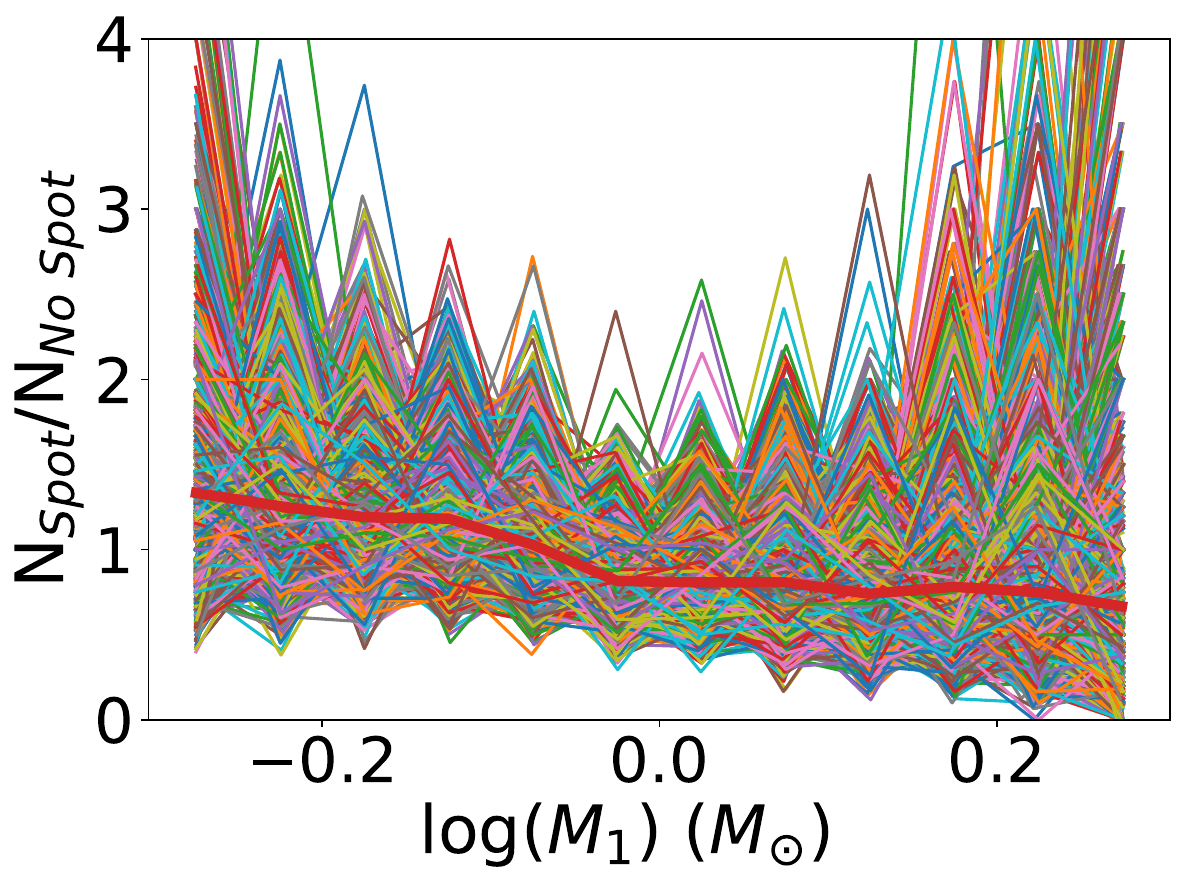}{0.48\textwidth}{\raisebox{1.5ex}{(2)}}
         }
\caption{The distributions of $N_W/N_A$ (Panel 1) and $N_{\text{Spot}}/N_{\text{No Spot}}$ (Panel 2) as functions of artificially perturbed primary masses. A total of 5,000 simulated datasets were generated using the Monte Carlo method, each shown as a colored line. The thick red line represents the median of all distributions. \label{fig:WA_distribution_by_false_mass}}
\end{figure*}

Figure \ref{fig:WA_distribution_by_false_mass} presents the distributions of $N_W/N_A$ and $N_{\text{Spot}}/N_{\text{No Spot}}$ as functions of the simulated primary star mass, generated from the Monte Carlo simulation described in Section \ref{sec:uncertainties_and_impact}.

In total, 5,000 sets of artificial primary masses were created by applying random relative deviations to the original values. The deviations follow the same distribution as the observed deviations shown in Figure \ref{fig:Compar_stellar_model_and_orbital_dynamical}. For each of the 5,000 perturbed datasets, we recalculated the distributions as to Panels 1 and 4 of Figure \ref{fig:dist_WA_spots}. Each colored line in Figure \ref{fig:WA_distribution_by_false_mass} represents one of these recalculated distributions.

The distribution of $N_W/N_A$ exhibits a clear decreasing trend. Although $N_{\text{Spot}}/N_{\text{No Spot}}$ shows large fluctuations at both ends due to the small sample size in those bins, the overall downward trend remains valid.


\section{\texorpdfstring{The Relationship Between $N_W/N_A$, $N_{\text{Spot}}/N_{\text{No Spot}}$, and Secondary Star Mass $\log{M_2}$}{The Relationship Between NW/NA, NSpot/NNo Spot, and Secondary Star Mass log(M2)}} \label{appendix:relation_for_m2}

\begin{figure*}[h]
\gridline{\fig{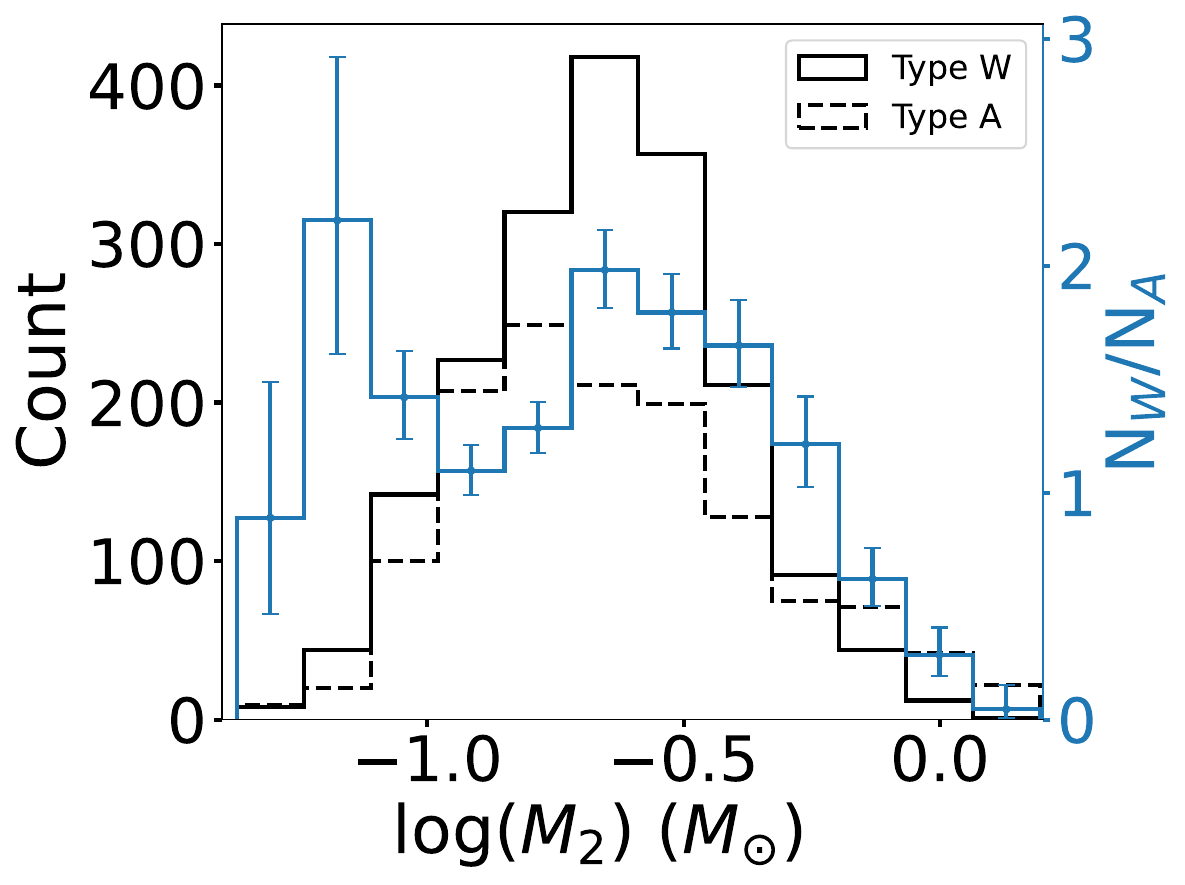}{0.49\textwidth}{\raisebox{1.5ex}{(1)}}
          \fig{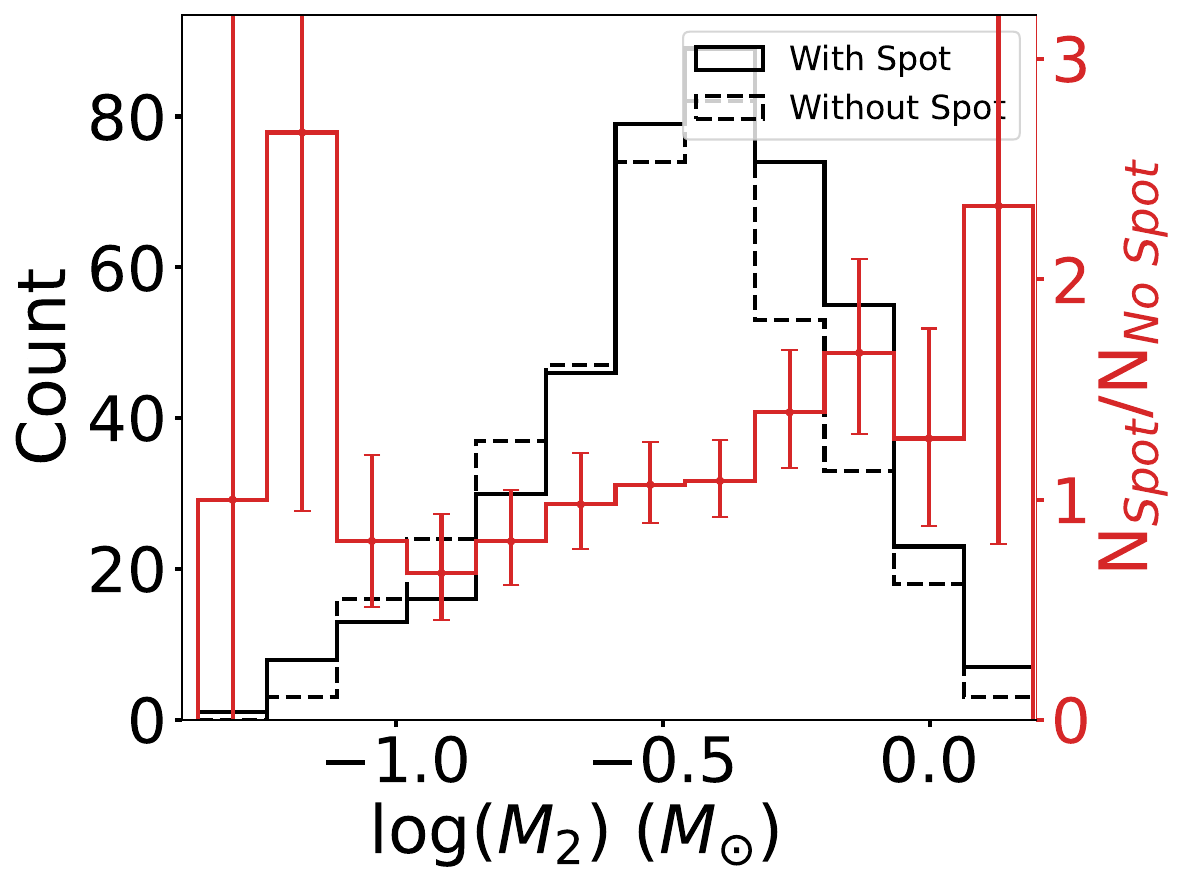}{0.49\textwidth}{\raisebox{1.5ex}{(2)}}
         }
\caption{Same as Panels 1 and 4 in Figure \ref{fig:dist_WA_spots}, but as a function of secondary mass. \label{fig:dist_WA_spots_for_M2}}
\end{figure*}

Similar to Panels 1 and 4 in Figure \ref{fig:dist_WA_spots}, the distribution of these ratios as a function of secondary mass is shown in Figure \ref{fig:dist_WA_spots_for_M2}. The results indicate that neither $N_W/N_A$ nor $N_{\text{Spot}}/N_{\text{No Spot}}$ exhibits a general correlation with the secondary mass, and their variation trends do not align with each other. This suggests that the secondary mass plays a minimal role in determining the contact binary type, further reinforcing the notion that the primary mass is more closely linked to the classification. Additionally, this provides indirect support for the hypothesis that magnetic activities on the primary star—primarily in the form of starspots—is the dominant factor driving the W-type phenomenon.

\begin{acknowledgments}
This work is supported by the Science Foundation of Yunnan Province （grant Nos. 202503AP140013, 202501AS070055, 202401AS070046, 202401AW070004), and the International Partnership Program of Chinese Academy of Sciences (No. 020GJHZ2023030GC), the China Manned Space Program with grant no. CMS-CSST-2025-A16, and the Yunnan Revitalization Talent Support Program. X.Z.L acknowledges the support of the Anhui Provincial Natural Science Foundation (2308085QA35). This research has made use of the SIMBAD database, operated at CDS, Strasbourg, France. This work has made use of data from the European Space Agency (ESA) mission {\it Gaia} (\url{https://www.cosmos.esa.int/Gaia}), processed by the {\it Gaia} Data Processing and Analysis Consortium (DPAC, \url{https://www.cosmos.esa.int/web/Gaia/dpac/consortium}). Funding for the DPAC has been provided by national institutions, in particular the institutions participating in the {\it Gaia} Multilateral Agreement.
Guoshoujing Telescope (the Large Sky Area Multi-Object Fiber Spectroscopic Telescope \textit{LAMOST}) is a National Major Scientific Project built by the Chinese Academy of Sciences. Funding for the project has been provided by the National Development and Reform Commission. \textit{LAMOST} is operated and managed by the National Astronomical Observatories, Chinese Academy of Sciences.
\end{acknowledgments}

\vspace{5mm}
\facilities{Gaia, Lamost}


\bibliography{sample701}{}
\bibliographystyle{aasjournalv7}



\end{CJK}
\end{document}